\documentclass[fleqn,10pt]{wlscirep}
\usepackage[utf8]{inputenc}
\usepackage[T1]{fontenc}
\usepackage{bm}
\usepackage{siunitx}
\title{Pruning random resistive memory for optimizing analogue AI}

\author[1,2,4,8,9,$^{\dagger}$]{Yi Li}
\author[1,2,3,9,$^{\dagger}$]{Songqi Wang}
\author[1,3,9]{Yaping Zhao}
\author[1,2,3,9]{Shaocong Wang}
\author[2,4,8]{Woyu Zhang}
\author[1,9]{Yangu He}
\author[1,2,3,9]{Ning Lin}
\author[1,9]{Binbin Cui}
\author[1,9]{Xi Chen}
\author[1,9]{Shiming Zhang}
\author[5]{Hao Jiang}
\author[6]{Peng Lin}
\author[5]{Xumeng Zhang}
\author[1]{Xiaojuan Qi}
\author[1,2,9,*]{Zhongrui Wang}
\author[2,4,8,*]{Xiaoxin Xu}
\author[2,4,8,*]{Dashan Shang}
\author[2,5]{Qi Liu}
\author[3,7]{Kwang-Ting Cheng}
\author[2,5]{Ming Liu}
\affil[1]{Department of Electrical and Electronic Engineering, the University of Hong Kong, Hong Kong, China}
\affil[2]{Laboratory of Microelectronic Devices \& Integrated Technology, Institute of Microelectronics, Chinese Academy of Sciences, Beijing 100029, China}
\affil[3]{ACCESS – AI Chip Center for Emerging Smart Systems, InnoHK Centers, Hong Kong Science Park, Hong Kong, China}
\affil[4]{State Key Lab of Fabrication Technologies for Integrated Circuits, Institute of Microelectronics, Chinese Academy of Sciences, Beijing 100029, China}
\affil[5]{Frontier Institute of Chip and System, Fudan University, Shanghai 200433, China}
\affil[6]{College of Computer Science and Technology, Zhejiang University, Zhejiang, 310027, China}
\affil[7]{Department of Electronic and Computer Engineering, the Hong Kong University of Science and Technology, Hong Kong, China}
\affil[8]{University of Chinese Academy of Sciences, Beijing 100049, China}
\affil[9]{Institute of Mind, the University of Hong Kong, Hong Kong, China}

\affil[$^{\dagger}$]{These authors contributed equally.}
\affil[*]{e-mail: zrwang@eee.hku.hk; xuxiaoxin@ime.ac.cn; shangdashan@ime.ac.cn}

\begin{abstract}

The rapid advancement of artificial intelligence (AI) has been marked by the large language models exhibiting human-like intelligence. However, these models also present unprecedented challenges to energy consumption and environmental sustainability, which are exacerbated by the increasing number of users and the development of even larger models. One promising solution is to revisit analogue computing, a technique that predates digital computing and exploits emerging analogue electronic devices, such as resistive memory, which features in-memory computing, high scalability, and nonvolatility that addresses the von Neumann bottleneck, slowdown of Moore's law, and volatile DRAM of conventional digital hardware. However, analogue computing still faces the same challenges as before: programming nonidealities and expensive programming due to the underlying devices physics. Therefore, leveraging the efficiency advantage while mitigating the programming disadvantage of analogue computing with resistive memory is a major open problem in AI hardware and electronics communities.
Here, we report a universal solution, software-hardware co-design using structural plasticity-inspired edge pruning to optimize the topology of a randomly weighted analogue resistive memory neural network. Software-wise, the topology of a randomly weighted neural network is optimized by pruning connections rather than precisely tuning resistive memory weights. Hardware-wise, we reveal the physical origin of the programming stochasticity using transmission electron microscopy, which is leveraged for large-scale and low-cost implementation of an overparameterized random neural network containing high-performance sub-networks.
We implemented the co-design on a 40nm 256K resistive memory macro, observing 17.3\% and 19.9\% accuracy improvements in image and audio classification on FashionMNIST and Spoken digits datasets, as well as 9.8\% (2\%) improvement in PR (ROC) in image segmentation on DRIVE datasets, respectively. This is accompanied by 82.1\%, 51.2\%, and 99.8\% improvement in energy efficiency thanks to analogue in-memory computing.
By embracing the intrinsic stochasticity and in-memory computing, this work may solve the biggest obstacle of analogue computing systems and thus unleash their immense potential for next-generation AI hardware.

\end{abstract}
\begin{document}

\flushbottom
\maketitle

\thispagestyle{empty}

\section*{Introduction}

Recent advancements in artificial intelligence (AI), especially the emergence of multimodal large language models like GPT-4V~\cite{LanguageModels,Radford2018ImprovingLU}, have revolutionized natural language and image processing and exhibited human-like intelligence~\cite{VaswaniSPUJGKP17,BERT,vit,HuggingFace}. The trend of developing and deploying ever-larger and more sophisticated models requires substantial computational power, which entails significant challenges in energy consumption and environmental sustainability~\cite {Energy}. As these models become more prevalent and indispensable, the concerns regarding their energy demand and ecological implications are exacerbated~\cite {JMLR}.

Revisiting analogue computing~\cite{Hogan}---a technique predating digital computing---presents a promising solution. By harnessing emerging analogue devices, such as resistive memory~\cite{1083337,strukov2008missing,huh2020memristors,Tsuyoshi3,Tsuyoshi2}, analogue computing directly processes informative analogue signals, bolstering energy efficiency in three ways~\cite{RN204,RN2575,RN2576,RN2584,RN2605,RN2577,RN2587,RN2592,RN2591,RN2559,RN2560,RN2556,RN2553,RN2589,RN2549}. First, digital computers feature physically separated memory and processing units, leading to significant time and energy overheads due to massive and frequent data shuttling, the so-called von Neumann bottleneck~\cite{RN2606,RN2566,RN2555,RN2550,RN2558,RN2570,RN2563,RN2564,RN2551,RN2548,CHEN2020264}. In contrast, analogue resistive memory collocate storage and processing within the same physical device~\cite{xi2020memory,mckee2004reflections,kuroda2001cmos,horowitz20141,Valovarticle,Valovarticle2}. Second, as the size of complementary metal-oxide-semiconductor (CMOS) approaches its physical limit, Moore’s law, which has fueled the development of transistors for years, is slowing down~\cite{theis2017end,schaller1997moore,shalf2015computing,shalf2020future}. Unlike CMOS, analogue resistive memory can be highly scalable and stackable~\cite{joshi2020accurate,wong2010phase,koelmans2015projected,soni2014giant,xi2017giant,wen2013ferroelectric}. Third, the main memory of digital computers, dynamic random-access memory (DRAM) is volatile, whereas analogue resistive memory can retain data without electricity~\cite{RN2578,RN201,RN2593,RN2585,RN2579,RN35,li2022mixed,RN2574,RN2552,zhang2022few,ambrogio2018equivalent,RN2568,Tsuyoshi}.

Nonetheless, emerging analogue computing systems persistently encounter long-standing challenges: programming nonidealities and expensive programming. Analogue resistive memory exhibit stochasticity and nonlinearity during programming~\cite{RN2554,RN2546,ambrogio2014statistical,dalgaty2021situ,burr2015experimental,sebastian2015collective,RN2521,RN2608,li2023adc,RN202,RN2603}. Furthermore, the energy and time expenditures associated with programming analogue resistive memory are significantly larger than those of digital devices~\cite{RN2566,RN2569,chih202116,wong2012metal,valov2013nanobatteries,RN2573}. Therefore, leveraging the efficiency advantage while mitigating the programming disadvantage of analogue computing with emerging resistive memory is a major open problem in AI hardware and electronics communities.

To tackle the aforementioned issues, we present a novel analogue computing paradigm, the software-hardware co-designed edge pruning topology optimization for randomly weighted resistive memory neural networks. Drawing inspiration from the clipped Hebbian-rule-based structural plasticity~\cite{marcus1989stability}, this co-design mimics the human brain's postnatal process of overproduction and consolidation of functional synapses, while eliminating redundant ones after a prolonged learning period~\cite{sakai2020synaptic,sretavan1984prenatal}. In contrast to conventional weight optimization methods that precisely adjust resistive memory conductance~\cite{RN2575,RN2576,RN2605,RN2577,RN2580,RN2553,RN35,RN2604,RN2547,RN2545,RN2569}, the edge pruning topology optimization directly engineers the topology of a randomly initialized neural network by "turning off" insignificant resistive memory weights and maintaining the conductance of the rest. Furthermore, we capitalize on the intrinsic electroforming stochasticity of the resistive memory to generate large-scale, low-cost hardware random weights, transforming the programming stochasticity into a benefit. According to Ramanujan et al.’s theory~\cite{ramanujan2020s}, at least one sub-network pruned from a neural network with sufficient random weights can achieve competitive accuracy compared to the original network with well-optimized weights, so we physically reset (set) resistive memory to prune (reinstate) the edge of the neural network. Since precise resistive memory programming is not required, the edge pruning topology optimization is not only intrinsically robust to analogue device non-idealities but also gets rid of the tedious conductance tuning and verification in conventional weight optimization. 

Here, we validated our co-design on three representative tasks, image and audio classification, and image segmentation, on a hybrid analogue-digital system with a 40nm 256K resistive memory-based analogue in-memory computing core. We achieved 17.3\% and 19.9\% accuracy improvements on the FashionMNIST and Spoken Digit datasets while reducing weight updates by 99.94\% and 99.93\% compared to the hardware-in-loop weight optimization. The associated inference energy for a single image and audio is reduced by 82.1\% and 51.2\% compared to state-of-the-art graphics processing units (GPUs), respectively, due to analogue in-memory computing. In addition, we simulate segmenting the DRIVE dataset with a U-Net on our co-design. The area under simulated precision-recall (PR) curves (AUC) and receiver operating characteristic (ROC) on the DRIVE dataset are 0.91 and 0.97, respectively, corresponding to 9.8\% and 2\% performance improvements compared to GPU. Meanwhile, the inference energy is reduced by 99.8\% owing to the in-memory computing and sparse weights. This work provides a universal solution for AI using analogue computing with emerging resistive memory, demonstrating the great potential for the advancement of future AI hardware with brain-like energy efficiency.

\begin{figure}[!t]
\centering
\includegraphics[width=0.9\linewidth]{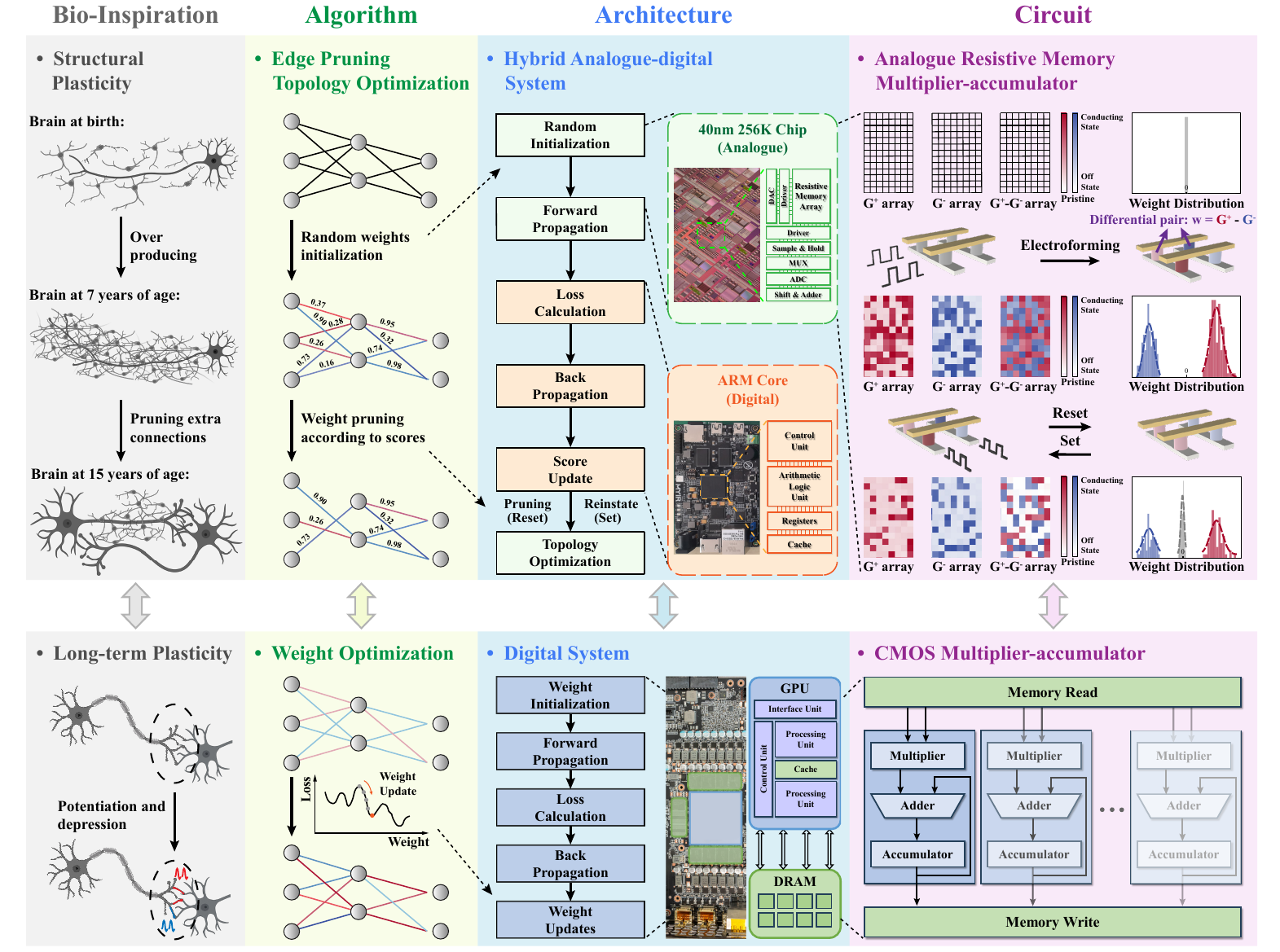}
\caption{\textbf{Software-hardware co-designed edge pruning topology optimization.} 
Comparison between the software-hardware co-design (resistive memory-based edge pruning topology optimization) and weight optimization on conventional graphic processing units (GPU) across bio-inspiration, algorithm, architectural, and circuit levels. At the bio-inspiration level, our approach (upper panel) is influenced by the observed pruning of overproduced neurons in children's brains through repeated learning and experiences, leaving only efficient connections. This contrasts with the conventional approach inspired by the simple long-term plasticity (lower panel). At the algorithm level, the edge pruning topology optimization engineers the topology of a randomly weighted neural network (upper panel). Initially, an overparameterized neural network with dense connections is randomly generated for subsequent pruning optimization, as shown in the left panel. During training, redundant connections are pruned, yielding a sparse sub-network of valuable connections. This contrasts with the precise weight adjustment (lower panel). At the architecture level, the analogue core of the hybrid analogue-digital system minimizes data transfers between the processor and memory (upper panel) compared to the conventional memory-processor-separated digital system (lower panel), mitigating the von Neumann bottleneck and offering high energy efficiency. At the circuit level, resistive memory not only implements more efficient and parallel analogue matrix multiplication (upper panel) compared to the digital alternative (lower panel) but also implements the edge pruning topology optimization. The intrinsic stochasticity in electroforming resistive memory generates high-density and low-cost stochastic weights, as revealed by the differential conductance heatmap and histogram. During topology optimization, edge pruning is implemented by resetting corresponding resistive memory differential pairs to nearly zero conductance, whereas reinstatement is realized by setting the pruned cells back to the conducting state.}
\label{fig1}
\end{figure}

\section*{Software-hardware co-design: edge pruning topology optimization for a randomly weighted resistive memory neural network}
Fig.~\ref{fig1} schematically illustrates the software-hardware co-designed edge pruning topology optimization on a randomly weighted resistive memory neural network.

Software-wise, the bio-inspired edge pruning topology optimization mimics the human brain's postnatal process of overproduction and consolidation of functional synapses, while eliminating redundant ones after a prolonged learning period (upper left panel of Fig. \ref{fig1}). Different from conventional weight optimization methods which train neural networks by adjusting weights to minimize loss, the edge pruning topology optimization tailors the network architecture to learn an effective sub-network without weight tuning. This approach is based on the complimentary conjecture of the Lottery Ticket Hypothesis~\cite{frankle2018lottery,ramanujan2020s}, which suggests that pruned sub-networks from an overparameterized neural network can achieve competitive accuracy compared to the original neural network with well-optimized weights. As demonstrated in the second panel of Fig. \ref{fig1}, edge pruning topology optimization begins with random initialization of the resistive memory network using electroforming stochasticity. Each edge features a fixed random weight and a corresponding score that assesses the connection’s importance. During the forward pass, the weights with lower scores in each layer, regarded as redundant connections, are pruned to formulate the sub-network. On the backward pass, we update all the scores and selectively replace certain edges with others to optimize the topology of the sub-network and minimize the training error (see Methods).

Hardware-wise, the proposed edge pruning topology optimization is physically implemented on a hybrid analogue-digital computing system. This system comprises two parts: an analogue core built on the 40nm 256K resistive memory in-memory computing macro for implementing random weights, accelerating computation-expensive matrix multiplications and pruning redundant edges; and a digital core consisting of a Xilinx system-on-chip (SoC) (see Methods for details). As resistive memory collocates both weight storage and computing, the analogue core of the hybrid system effectively reduces the data shuffling between memory and processing units, offering better energy efficiency and parallelism compared to conventional digital hardware. Before training, the resistive memory array is first split into positive and negative conductance ($\textbf{G}^+$ and $\textbf{G}^-$) sub-arrays, which represents the weight matrix using the conductance difference between two arrays, as shown in the right panel of Fig. \ref{fig1}. As-deposited resistive memory cells are intrinsically insulating, resulting in a narrow weight distribution around zero conductance. Subsequently, $\textbf{G}^+$ and $\textbf{G}^-$ arrays are electroformed, yielding random and analogue hardware weights. The conductance distribution of resistive memory differential pairs follows a mixture of two quasi-normal distributions by leveraging the intrinsic stochasticity of electroforming as a natural source of randomness (see Supplementary Information Table 1 for the randomness studies), offering large-scale, low-cost, and true random weights for the overparameterized neural network. During training, if an edge is identified as an unnecessary connection that needs to be removed from the original neural network, the weight is physically pruned by resetting the corresponding resistive memory differential pair, annihilating the conductive channels and zeroing their conductance. Conversely, if a previously pruned connection proves important and needs to be reinstated, the resistive memory pair is set back to its conducting state, which recovers the conductive channel (see Methods for details of reset and set operations). After training, we froze the high-performing sub-network, and the conductance distribution shows a mixture of three quasi-normal distributions, where the pruned cells also reduce forward pass energy.

\begin{figure}[!t]
\centering
\includegraphics[width=0.9\linewidth]{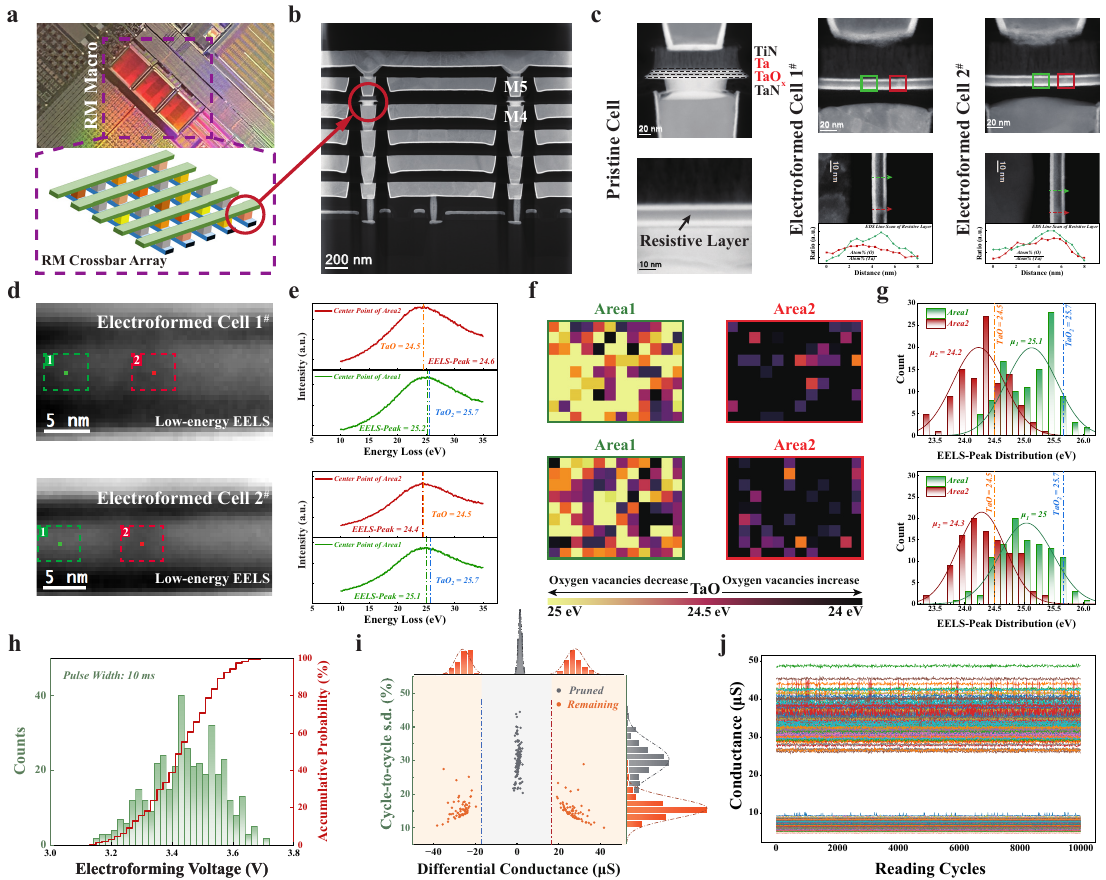}
\caption{\textbf{Physical and electrical characterization of the resistive memory (RM) electroforming stochasticity.}
\textbf{a,} Optical photo of the 40nm 256K resistive memory in-memory computing macro and the schematic of the resistive memory crossbar array.
\textbf{b,} Cross-sectional HAADF–STEM image of the resistive memory array, fabricated between the metal 4 and metal 5 layers using the back-end-of-line process. 
\textbf{c,} Cross-sectional HAADF-STEM images and EDS line profiles of electroformed resistive memory cells. The red (green) arrow is from the red-boxed (green-boxed) areas with (without) distinct structural differences.
\textbf{d,e,} EELS plane scans (energy range: 15eV-35eV, step: 0.05eV) and corresponding center point's low-loss spectra of two electroformed resistive memory cells, where the green (Area1, insulating) and red (Area2, conducting) regions correspond to the green and red boxes in \textbf{c}.
\textbf{f,} EELS low-loss peak maps of Area1 and Area2, where the region's colour changes from yellow to black as the concentration of oxygen vacancies increases. The geometries of the conductive paths in the two cells show significant differences.
\textbf{g,} The corresponding EELS-peak distributions in \textbf{f}, where the green and red curves refer to Area1 and Area2, respectively.
\textbf{h,} Histogram and accumulative probability of the electroforming voltages of a 20×20 resistive memory array, where a linear voltage sweep starting from 3V with a step of 0.05V is applied to each cell.
\textbf{i,} Joint distribution of the conductance and standard deviation of 128 randomly selected resistive memory differential pairs after 100 pruning and reinstating cycles. The pruned (remaining) pairs are represented by grey (orange) points. The probability densities of the conductance and standard deviation are displayed on the top and right histograms.
\textbf{j,} Data retention of the 128 randomly selected resistive memory cells over 10,000 reading cycles.
}
\label{fig2}
\end{figure}

\section*{Physical origin of resistive memory array programming stochasticity}
We first fabricated nanoscale resistive memory cells and electroformed them under identical conditions. The programmed cells are then sliced using a focused ion beam (FIB) before being examined by a high-resolution transmission electron microscope (TEM) to elucidate the microstructural origin of the programming stochasticity.

Fig. \ref{fig2}a shows the optical photo of the resistive memory in-memory computing macro. The monolithically integrated 256K macro features a crossbar structure, where resistive memory cells in each row share bottom electrodes and those in each column share top electrodes. These cells are integrated with CMOS on the 40nm standard logic platform between the metal 4 and metal 5 layers using the back-end-of-line process, as revealed by the high-angle annular dark-field (HAADF) scanning transmission electron microscope (STEM) (Fig. \ref{fig2}b).
 Fig. \ref{fig2}c shows distinct differences in the cross-sectional HAADF-STEM images of resistive memory cells with and without electroforming. In the pristine cell (left), the tantalum and tantalum oxide (Ta/TaO$_x$) resistive layer exhibits good uniformity (e.g. green-boxed area), while in electroformed cells (middle and right), several brighter contrast regions (e.g. red-boxed areas) are observed between the top and bottom electrodes, which exhibit clear structural differences and are likely to be conducting channels. To probe the composition, we carry out the energy-dispersive X-ray spectroscopy (EDS) line scan. Brighter regions (red arrow) reveal a lower ratio of oxygen to tantalum concentration than the dark region (green arrow), confirming the existence of conducting channels~\cite{RN2554,park2013situ,RN2546}. In addition, the red line profiles of the two cells are significantly different, which is consistent with the observed electroforming stochasticity in Fig. \ref{fig2}h,i. The corresponding low-energy electron energy loss spectroscopy (EELS) plane scans and center point's spectra (Fig. \ref{fig2}d,e) of electroformed cells further reveal the changes in oxygen-vacancy concentrations and valence between the dark region (Area1) and bright region (Area2). The losses in Area1 (25.2eV and 25.1eV) are close to the plasmon peak of the insulating TaO$_2$ phase (25.7eV), while the losses in Area2 (24.6eV and 24.4eV) are similar to the metallic TaO phase (24.5eV), reflecting the oxygen-vacancy migration during electroforming~\cite{park2013situ,li2017direct} and is consistent with EDS observations. Additionally, we visualize the shape of the conductive channels in the tested area by extracting the EELS low-loss peak maps to illustrate the oxygen-vacancy-rich conducting paths in the resistive layer (Fig. \ref{fig2}f). As expected, the conductive TaO$_x$ clusters span almost the entire Area2, while no fully-formed conducting channels are developed in Area1. The distinct patterns of conductive channels and the corresponding EELS-peak distributions (Fig. \ref{fig2}g) reveal the inherently stochastic nature of the electroforming due to the inevitable thin film inhomogeneity and random motion of oxygen ions, offering low-cost and highly scalable random conductance to physically implement randomly weights of neural networks.

The intrinsic stochasticity of electroforming will be harnessed to physically implement a randomly weighted neural network for the proposed edge pruning topology optimization. Fig. \ref{fig2}h illustrates the process of generating random weights using the resistive memory array. Firstly, we sample the electroforming voltage from a small-scale resistive memory sub-array, where the electroforming voltage is defined as the minimum voltage that switches cell resistance from $\sim$30M$\Omega$ to less than 300K$\Omega$. Based on the obtained accumulative probability, we then apply a uniform voltage with a 3.4V amplitude and 10ms width to the $\textbf{G}^+$ array at 120°C, generating a random conductance matrix where around half of the cells are successfully electroformed while the others remain insulating (sparsity equals 0.5). Finally, we electroform each cell in the $\textbf{G}^-$ array into the complementary state with respect to the $\textbf{G}^+$ array (see Supplementary Information Fig.3 for the impact of random weight distributions).
During training, the resistive memory weights are pruned by resetting the corresponding resistive memory differential pairs. As shown in Fig. \ref{fig2}i, analogue hardware weights follow a mixture of three quasi-normal distributions, and the mean conductance of the pruned differential pairs is around 0.07$\mu$S with small variation (see Supplementary Fig. 4-6 and Notes for robustness studies).
Fig. \ref{fig2}j shows the data retention of resistive memory cells with a small conductance fluctuation over 10,000 reading cycles, which reduces overfitting during training as to be discussed in the subsequent sections.

\begin{figure}[!t]
\centering
\includegraphics[width=0.9\linewidth]{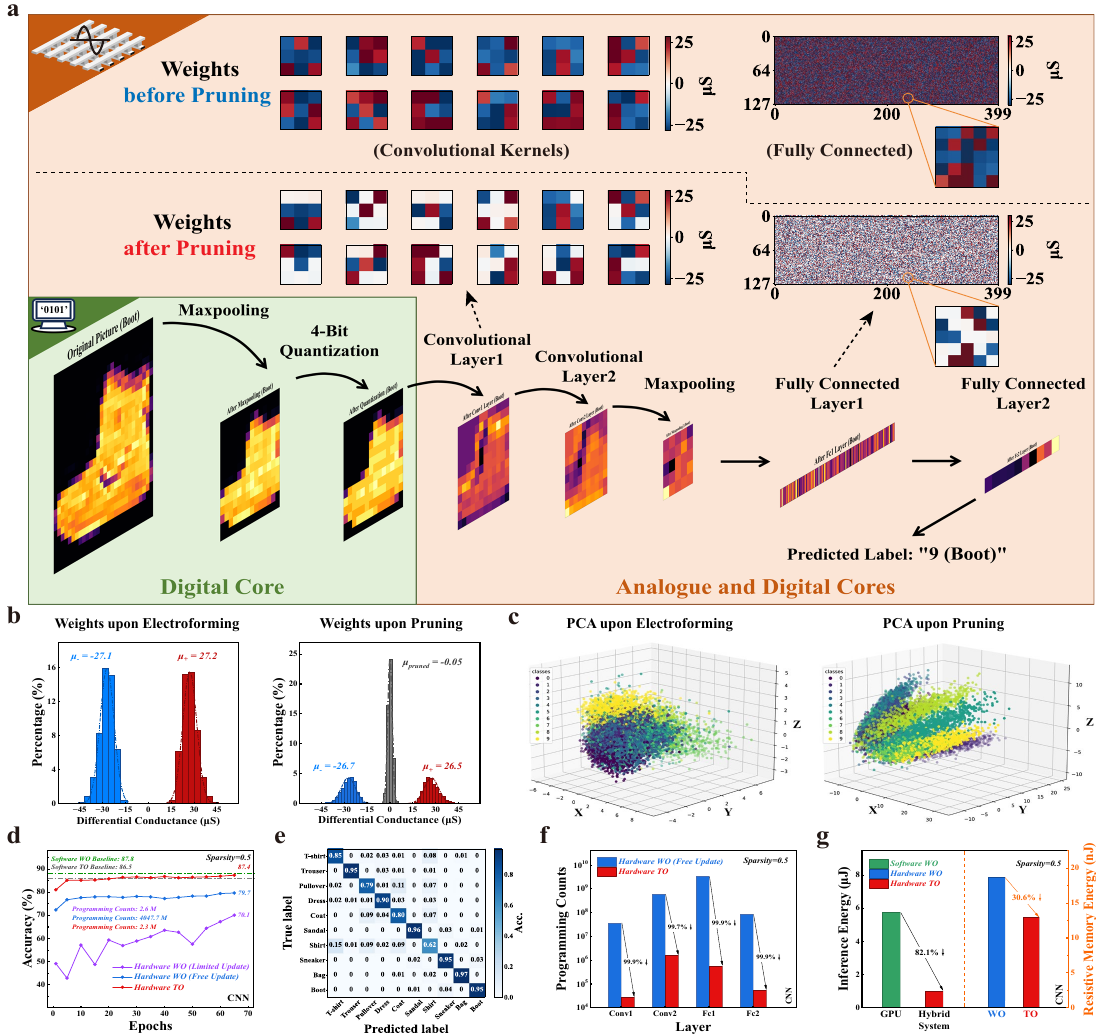}
\caption{\textbf{Experimental image classification for FashionMNIST dataset with the co-design.}
\textbf{a,} Schematic illustration of the forward pass in garment classification. Test images are pre-processed digitally and input to a randomly weighted four-layer CNN physically implemented on analogue random resistive memory (RM). The upper shows the measured hardware weights upon electroforming and topology optimization, where pruned cells are of nearly zero conductance (white) while the conductance of the remaining cells is fixed (blue and red).
\textbf{b,} Corresponding hardware weight distributions upon electroforming and edge pruning topology optimization, where a mixture of three quasi-normal distributions with significant sense margins is observed after pruning.
\textbf{c,} 3D PCA of the feature embedding upon electroforming and edge pruning topology optimization, where clear clusters are observed in the latter. 
\textbf{d,} Measured classification accuracy during 65 training epochs. The hardware edge pruning topology optimization (TO) achieves an accuracy of 87.4\%, a 7.7\% improvement over the hardware weight optimization (WO) with free updates, and a 0.4\% accuracy loss compared to the software weight optimization baseline. When restricting the programming counts (2.6 million) using a gradient threshold, the weight optimization with similar per resistive memory cell updates as topology optimization (2.3 million) suffers a significant 17.3\% accuracy drop.
\textbf{e,} Experimental confusion matrix, dominated by the diagonal elements.
\textbf{f,} Breakdown of the programming counts for hardware-in-the-loop weight optimization and topology optimization. The latter results in a 99.94\% reduction in weight updates compared to the former.
\textbf{g,} Comparison of inference energy of a single image. The left shows hybrid analogue-digital system saves 82.1\% of energy compared to GPU owing to in-memory computing. The right indicates that topology optimization reduces the resistive memory energy consumption by 30.6\% compared to weight optimization due to the sparsity.
}
\label{fig3}
\end{figure}

\section*{Image classification of FashionMNIST using the co-design}
We first assessed our co-design on a 4-layer Convolutional Neural Network (CNN)~\cite{kadam2020cnn}, a standard vision model, to classify the FashionMNIST~\cite{xiao2017fashion} dataset using the co-design. 

Fig. \ref{fig3}a presents example feature maps in garment classification with the edge pruning topology optimization and hybrid analogue-digital system. The FashionMNIST dataset, commonly used for image classification tasks, consists of 70,000 frontal images of ten garment categories. Test images are first down-sampled, quantified into 14×14 and 4-bit patterns, and then fed into a randomly weighted 4-layer CNN consisting of two convolutional and two fully connected layers (see Methods for the details of CNN). The CNN model is initialized with 62K random weights (124K randomly initialized resistive memory cells). The differential conductance heatmaps upon electroforming and edge pruning topology optimization are plotted in the upper part of the schematic. Fig. \ref{fig3}b shows the corresponding distribution. The conductance weights after electroforming follow a mixture of two quasi-normal distributions with mean values of -27.1$\mu$s and 27.2$\mu$s. After topology optimization, half of the resistive memory pairs (sparsity equals 0.5) are pruned, introducing another quasi-normal distribution with a mean of -0.05$\mu$s (see Supplementary Information Fig. 4 for CNN hyperparameter studies).
Fig. \ref{fig3}c visualizes the three-dimensional (3D) principal component analysis (PCA) of the embedded features for the classification head before and after edge pruning topology optimization, where each point is colour-coded to represent a specific category of garments. By optimizing the topology of the random weighted neural network, the originally overlapped embeddings of different garment categories transform into clearly segregated clusters, indicating the learned features are discriminative.

As shown in Fig. \ref{fig3}d, the experimental edge pruning topology optimization (hardware TO) achieves a classification accuracy of 87.4\% compared with 79.7\% of hardware weight optimization (hardware WO) with free updates, as the latter suffers from programming noise. This corresponds to a 0.4\% accuracy loss compared to the software weight optimization (software WO) baseline on GPU. The proposed edge pruning topology optimization also features better learning efficiency, revealing a 17.3\% accuracy margin over weight optimization with gradient threshold under the same budget of weight updates (see Methods for weight optimization details). Also, the hardware-in-loop edge pruning topology optimization improves the accuracy by 0.9\% compared to the software edge pruning optimization on GPU, primarily due to the mitigation of overfitting by cycle-to-cycle conductance variance. 
The high classification accuracy is further evidenced by the confusion matrix in Fig. \ref{fig3}e, dominant diagonal elements.
Fig. \ref{fig3}f compares training complexity, where the edge pruning topology optimization significantly reduces the number of hardware weight updates in the convolutional and fully connected layers by an average of 99.74\% and 99.98\% (For a fair comparison, each training protocol is initialized with similar-sized parameters and ended with similar accuracy, see Methods and Supplementary Table 2).
Fig. \ref{fig3}g compares the energy consumption required for classifying a single image between the hybrid analogue-digital system ($\sim$1.03$\mu$J) and the state-of-the-art GPU ($\sim$5.76$\mu$J), with the former benefiting from resistive in-memory computing (see Supplementary Information for details of energy consumption estimation). The corresponding forward pass energy with the resistive memory, shown in the right panel, corroborates the sparsity advantage of our edge pruning topology optimization ($\sim$12.94nJ) over the weight optimization ($\sim$18.65nJ), demonstrating a significant energy saving of 30.6\% (see Supplementary Table 6 for details of energy consumption estimation).

\begin{figure}[!t]
\centering
\includegraphics[width=0.9\linewidth]{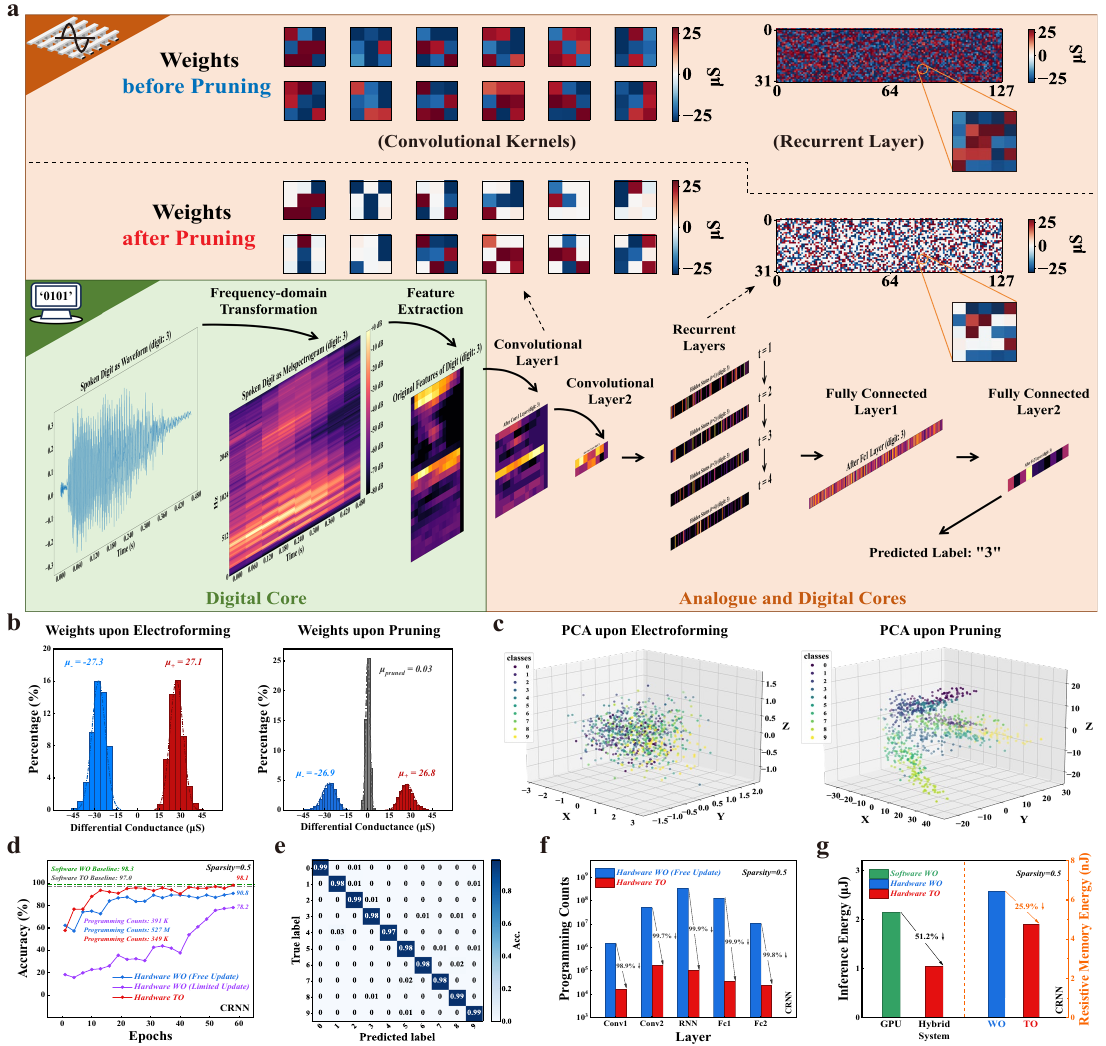}
\caption{\textbf{Experimental audio classification of Spoken Digit dataset with the co-design.} 
\textbf{a,} Schematic showing feature maps and selected weights in the audio classification of the Spoken Digit dataset. The raw audio is first converted to frequency-domain signals to generate a 23×15 feature map and then fed into a randomly weighted five-layer CRNN physically implemented on resistive memory (RM). During optimization, the pruned cells are reset to off-state (white pixels), while the rest cells' conductance remains fixed (blue and red pixels).
\textbf{b,} Corresponding hardware weight distributions upon electroforming and topology optimization. After pruning, nearly half of the cells are pruned, resulting in a mixture of three quasi-normal distributions.
\textbf{c,} 3D PCA of the feature distributions upon electroforming and topology optimization, where the latter shows more segregated clusters.
\textbf{d,} Measured audio classification accuracy during 60 training epochs. Hardware edge pruning topology optimization (TO) achieves a classification accuracy of 98.1\% compared with 90.8\% of hardware weight optimization (WO) with free updates, corresponding to a 0.2\% accuracy loss compared to the software weight optimization baseline. When restricting the number of programming (391 thousand) close to that of the topology optimization (349 thousand) by imposing the gradient threshold, the limited weight update optimization shows a 19.9\% accuracy drop.
\textbf{e,} Experimental confusion matrix with large diagonal elements.
\textbf{f,} Breakdown of the programming counts for hardware-in-the-loop weight and topology optimization. The latter features a reduction of 99.93\% programming counts compared to the weight optimization with the free update. 
\textbf{g,} Comparison of inference energy of a single image. The left shows hybrid analogue-digital system saves 51.2\% of energy compared to GPU due to in-memory computing. The right indicates that our proposed topology optimization has shown a 25.9\% reduction in the resistive memory energy consumption compared to the weight optimization due to sparsity.
}
\label{fig4}
\end{figure}

\section*{Audio classification for Spoken Digit with the co-design}
In our second experiment, our co-design classifies audios using a convolutional recurrent neural network (CRNN)~\cite{choi2017convolutional}, a classical model for extracting spatial and temporal features in audios through convolutional and recurrent layers, respectively. We use the Spoken Digit~\cite{zohar_jackson_2018_1342401} dataset, comprising a total of 3,000 recordings from 6 speakers at 8KHz. 

Fig. \ref{fig4}a illustrates the experimental forward pass feature maps using the hybrid analogue-digital system. The spoken digits undergo frequency-domain transformations and are then converted into 23×15 acoustic feature maps. Subsequently, the features are input into a randomly weighted 5-layer CRNN, which comprises 2 convolutional layers, 1 recurrent layer, and 2 fully connected layers (see Methods for the details of CRNN). The CRNN model contains 68.5K stochastic weights and is physically realized through 137K randomly initialized resistive memory cells. The hardware weight heatmaps upon electroforming and edge pruning are depicted in the upper part of the schematic. The hardware weights of the convolutional, recurrent, and fully connected layers are random after electroforming, following a mixture of two quasi-normal distributions with mean values of -27.3$\mu$s and 27.1$\mu$s (Fig. \ref{fig4}b, left panel). During edge pruning topology optimization, half of the resistive memory differential pairs (sparsity equals 0.5) are reset to nearly zero conductance, while the conductance of the remaining cells is fixed, showing a mixture of three quasi-normal distributions (Fig. \ref{fig4}b, right panel).
Fig. \ref{fig4}c presents the visualization of the measured 3D PCA of the embeddings for the classification head upon electroforming and edge pruning topology optimization. Each point is colour-coded to represent a specific category of spoken digits. Like the image classification, the initially overlapped embeddings of different spoken digits are clearly clustered after edge pruning topology optimization, signifying the discriminative features.

As shown in Fig. \ref{fig4}d, the experimental edge pruning topology optimization achieves a classification accuracy of 98.1\%, compared to 90.8\% of hardware-in-loop weight optimization, as the latter is prone to programming stochasticity. This results in a slight accuracy loss of 0.2\% compared to the software weight optimization baseline. Similarly to the previous image classification task, hardware-in-loop edge pruning topology optimization effectively prevents overfitting by exploiting the cycle-to-cycle conductance variation (see Supplementary Information Fig. 5 for CRNN hyperparameter studies).
The classification performance is corroborated by the confusion matrix in Fig. \ref{fig4}e with large diagonal elements. 
Fig. \ref{fig4}f highlights that edge pruning topology optimization significantly minimizes the number of hardware weight updates compared to weight optimization without update limits, resulting in an average reduction of 99.64\%, 99.97\%, and 99.96\% programming counts in the convolutional, recurrent, and fully connected layers, respectively.
The estimated energy consumption for inferring a single spoken digit is shown in Fig. \ref{fig4}g. In the left panel, the hybrid analogue-digital system consumes $\sim$1.05$\mu$J, while the GPU consumes $\sim$2.15$\mu$J, resulting in a 51.2\% inference energy saving. The resistive memory energy consumption in the right panel reveals the sparsity benefit associated with topology optimization ($\sim$4.79nJ) compared to weight optimization ($\sim$6.46nJ), or a 25.9\% decrease.

\begin{figure}[!t]
\centering
\includegraphics[width=0.9\linewidth]{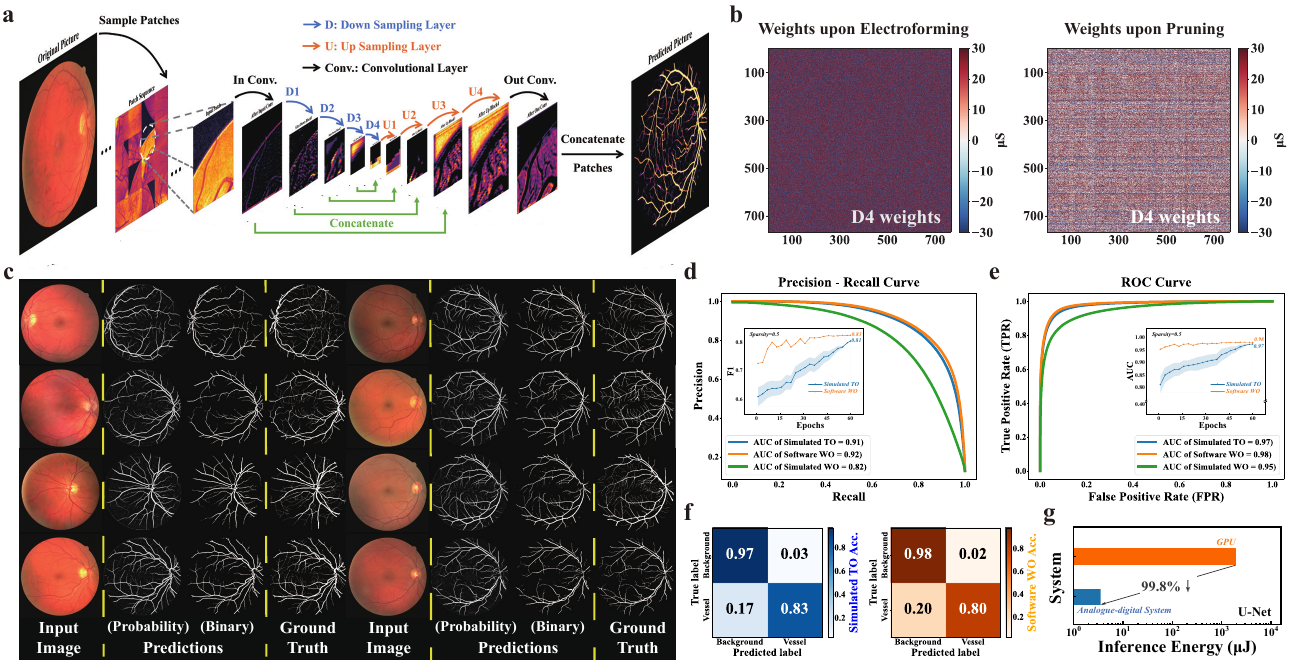}
\caption{\textbf{Simulated image segmentation for the DRIVE dataset with the co-design.} 
\textbf{a,} Illustration of the forward pass in segmenting the DRIVE dataset using the simulated co-design. The 584×565 blood vessel image is first divided into 96×96 patches. These patches are then segmented by the U-Net with random weights sampled from the conducting state resistive memory differential pair conductance distribution. Finally, the output patches are concatenated to produce the complete segmented image.
\textbf{b,} 768×768 random weights of the D4 layer upon electroforming and edge pruning. The pruned weights are sampled from the off-state resistive memory differential pair distribution (white pixels), while the remaining weights are sampled from the conducting state distribution (blue and red pixels).
\textbf{c,} Segmentation inputs (left), simulated probability and binary predictions of blood vessels (middle), and ground truth (right). The predictions are close to the ground truth.
\textbf{d,} Comparison of PR curves between software and simulated weight optimization (WO) as well as simulated edge pruning topology optimization (TO). The latter parallels the software weight optimization and outperforms the simulated weight optimization. The inset shows corresponding F1 curves.
\textbf{e,} Comparison of ROC curves between software and simulated weight optimization as well as simulated topology optimization. The inset shows the corresponding AUC curves.
\textbf{f,} Confusion matrix of the simulated pixel-wise blood vessel classification, dominated by the diagonal elements.
\textbf{g,} Comparison of estimated inference energy consumption. Our hybrid system demonstrates a 99.8\% energy reduction compared to software weight optimization on GPU due to in-memory computing.
}
\label{fig5}
\end{figure}

\section*{Image segmentation of DRIVE with the co-design}
In addition to classification, we simulate edge pruning optimizing a U-Net~\cite{huang2020unet,zhuang2018laddernet} for biomedical image segmentation using the Digital Retinal Images for Vessel Extraction (DRIVE)~\cite{staal2004ridge} dataset. This dataset contains 40 565×584 retinal images for segmenting and extracting blood vessels from the ocular background, including 7 normal pathology cases. Fig. \ref{fig5}a illustrates the forward pass in the U-Net (see Methods for the details of U-Net). The 584×565 blood vessel image is first divided into 96×96 patches before being input into the U-Net consisting of 2 convolutional layers (In and Out Conv.), 4 contracting down-sampling layers (D1-D4), and 4 expansive up-sampling layers (U1-U4). The U-Net produces segmented patches of the same size as the input patch, which are then concatenated as the output.
Fig. \ref{fig5}b shows the 768×768 simulated weights of the D4 layer upon electroforming and pruning. The former are sampled from the measured conductance distribution of electroformed resistive memory differential pairs. After topology optimization, half of the weights are pruned according to a pre-defined sparsity of 0.5 (white pixels).
The simulated segmentation results are shown in Fig. \ref{fig5}c, displaying the simulated probability and binary predictions as well as the ground truth, arranged from left to right. The simulated predictions closely resemble the ground truth, revealing a clear vessel structure and capillaries from the background (see Supplementary Information Fig. 6 for U-Net hyperparameter studies).
Fig. \ref{fig5}d and Fig. \ref{fig5}e compare the precision-recall (PR) and receiver operating characteristic (ROC) of the edge pruning and weight optimization on both the simulated hybrid analogue-digital system and software. The simulated edge pruning topology optimization displays an AUC of 0.91 and 0.97 for the PR and ROC curves, respectively, with a minor performance loss of 0.01 compared to the software weight optimization baseline, corresponding to 9.8\% and 2\% improvements compared to the simulated weight optimization thanks to its robustness against programming stochasticity. The insets show the corresponding F1-score (F1) and AUC curves of the simulated edge pruning optimization, which gradually approach the software weight optimization baselines during the training course.
Fig. \ref{fig5}f presents the confusion matrix, with the simulated U-Net achieving 97\% (83\%) accuracy in the background (vessel) pixel classification using our co-design, closely matching the 98\% (80\%) accuracy of the software weight optimization baseline (see Methods for PR, AUC, F1, and accuracy estimations).
Fig. \ref{fig5}g shows the estimated energy consumption for inferring a single image. The energy consumption for the resistive memory-based hybrid system is approximately 3.5$\mu$J while that of the state-of-the-art GPU is approximately 1913.6$\mu$J, resulting in a 99.8\% energy saving thanks to the highly efficient resistive memory based in-memory computing.

\section*{Discussion}
In this work, a software-hardware co-design, edge pruning topology optimization for randomly weighted resistive memory neural networks, is developed to address the challenges of AI implementation using analogue computing with emerging resistive memory. Hardware-wise, we embrace the intrinsic stochasticity of the resistive memory electroforming to produce large-scale, low-cost hardware random weights and directly optimize the topology of the neural network via reset operation to avoid precise conductance tuning, providing a time-energy efficient and robust way to harvest the analogue in-memory computing advantages for AI. Software-wise, the edge pruning topology optimization not only benefits from the true random weights offered by resistive memory arrays for initializing an overparameterized randomly weighted neural network but also significantly reduces the programming overhead of neural networks by pruning the redundant connections. This novel co-design may solve analogue computing's biggest obstacles of programming stochasticity and cost, laying the foundation for the next generation of AI hardware with brain-like energy efficiency.

\section*{Methods}

\subsection*{Fabrication of Resistive Memory Chip}

Under the 40nm technology node, the fabricated resistive memory chip features a 512×512 crossbar array, with resistive memory cells constructed between the metal 4 and metal 5 layers using the backend-of-line process. These cells comprise bottom and top electrodes (BE and TE) and a transition-metal oxide dielectric layer. The BE via, which possesses a diameter of 60nm, undergoes patterning through photo-lithography and etching. It is then filled with TaN via physical vapour deposition, followed by an overlay of a 10nm TaN buffer layer. Subsequently, a 5nm Ta layer is deposited and oxidized, resulting in the creation of an 8nm TaO$_x$ dielectric layer. The TE is formed by sequentially depositing 3nm Ta and 40nm TiN through physical vapour deposition. Upon completion of fabrication, the remaining interconnection metals are deposited using the standard logic process. The resistive memory cells in the same row share BE connections, while those in the same column share TE connections. After undergoing a 30-minute post-annealing process at 400°C in a vacuum, the 40nm resistive memory chip demonstrates exceptional performance, exhibiting high yield and robust endurance characteristics. (see Supplementary Information for resistive memory properties).

\subsection*{Hybrid Analog–Digital Hardware System}

The hybrid analogue-digital hardware system consists of a 40nm resistive memory chip and a Xilinx ZYNQ system-on-chip (SoC), which includes a field-programmable gate array (FPGA) and advanced RISC machines (ARM) processor integrated on a printed circuit board (PCB). The resistive memory chip operates in three primary modes according to the edge pruning topology optimization: electroform mode for generating random conductance weights, reset mode for pruning selected weights, and multiplication mode for vector-matrix products. The electroform mode triggers a dielectric breakdown in resistive memory arrays and forms random conductance matrices, with All source lines (SLs) biased to a fixed programming voltage sourced by an eight-channel digital-to-analogue converter (DAC, DAC80508, Texas Instruments) with 16-bit resolution, while bit lines (BLs) are grounded and word lines (WLs) are biased by the DAC to impose compliance current to cells and prevent hard breakdown. The SL voltage amplitude and width modulate the post-breakdown conductance distribution and sparsity. The reset mode restores a resistive memory cell back to its off-state, with the selected BL biased by the DAC, while the selected SL is grounded and the rest of the SLs float. For multiplication mode, a 4-channel analogue multiplexer (CD4051B, Texas Instruments) with an 8-bit shift register (SN74HC595, Texas Instruments) applies a DC voltage to the BLs of the resistive memory chip. During each training step, the resistive memory chip is read, and the multiplication values carried by the current from the SLs are converted to voltages using trans-impedance amplifiers (OPA4322-Q1, Texas Instruments) and analogue-to-digital converters (ADS8324, Texas Instruments, 14-bit resolution). These results are then sent to the Xilinx SoC for further processing. The FPGA contains logic that drives resistive memory and exchanges data with the ARM processor using direct double-data rate memory access via the direct memory access control unit. Additionally, the FPGA implements some neural network functions in hardware, such as activation and pooling.

\subsection*{Multi-Bit vector-matrix multiplications}

To execute vector-matrix multiplication, the initial stage entails converting the analogue input vector into an m-bit binary vector. In this process, each element is digitized as a binary number with m bits, where m equals 4 for CNN, 3 for CRNN, and 6 for U-Net. Consequently, analogue multiplication is approximated as a series of m times multiplications using binary input vectors corresponding to different significance levels. In each multiplication step, a row is either biased to a small fixed voltage (e.g., 0.1V) when it receives a '1' bit, while it is grounded when it receives a '0' bit. The resulting output currents from all columns are then sequentially collected via the column multiplexer. Finally, these collected currents undergo multiplication by their respective significance values and are aggregated within the digital domain.

\subsection*{Reset and Set Operations}

\subsubsection*{Edge Pruning Topology Optimization}

The weight pruning processes are physically realized by rendering corresponding resistive memory differential pairs in off-state using the reset operation, while the reinstatement of these weights is achieved by bringing the resistive memory cells back to their conducting states through the set operation. Here, the set operation is realized by applying identical pulses with a 3.3V amplitude and 300ns width to the bit line of the resistive memory array, recovering the pruned cells to the conducting state to reinstate into the sub-network. The reset operation is realized by applying identical pulses with a 2.6V amplitude and 400ns width to the source line of the resistive memory array, annihilating the conductive path. Importantly, it is noteworthy that there exists a significant margin between the off-state and the conducting state. Therefore, it is unnecessary to program the cells into precise conductance values.

\subsubsection*{Weight Optimization}

Identical pulses with a 1.5V amplitude and 500ns width are used to program resistive memory cells. The transistor's gate voltage limits the programming current through the word line to control conductance distribution. Each cell in differential resistive memory pairs is programmed to the target conductance using the closed-loop writing method~\cite{RN2545}. The following write verification ensures an approximately 10\% conductance error margin.

\subsection*{Edge Pruning Topology Optimization Algorithm}

The edge pruning topology optimization algorithm can be divided into forward and backward steps. Initially, a randomly weighted neural network is initialized via the analogue resistive memory chip, where the weight value also serves as the score for each edge in the network. On the forward pass, the sub-network is chosen by pruning the hardware weights with bottom-k\% lowest scores in each layer according to the pre-designed sparsity (i.e., sparsity equals 0.5 corresponding to 50\% pruned weights). Inputs are then fed into the sub-network for forward propagation and loss evaluation. On the backward pass, the general digital processor calculates the loss function's gradients to update the edges' scores, while fixing the weight values. The score is updated using the straight-through gradient estimator~\cite{bengio2013estimating,ramanujan2020s}:
\begin{equation}
\Tilde{s}_{ij} = {s_{ij} - \eta \times \frac{\partial{L}}{\partial{I_j}} \times w_{ij}Z_i}
\label{eq:straight-through gradient},
\end{equation}
where $s_{ij}$, $\Tilde{s}_{ij}$, $\eta$, $\frac{\partial{L}}{\partial{I_j}}$ and $w_{ij}Z_i$ denote the edge score between the hidden node i and j before and after the update, the learning rate, the partial derivative of loss ($L$) with respect to the input of node j ($I_j$) and the weighted output of node i, respectively. These processes are repeated until a well-performed sub-network is selected from the randomly initialized neural network.

\subsection*{Threshold Learning Rule}

To improve the energy and time efficiency during the training process, we follow Yao's work~\cite{RN2545} using the same threshold learning rule to reduce the unnecessary programming of resistive memory:

\subsubsection*{Edge Pruning Topology Optimization}

\begin{equation}
{\Delta}s=\begin{cases}
 {\Delta}s& \text{ if } |{\Delta}s|\ge {T_s(t)} \\
 0& \text{ } otherwise
\end{cases}
\label{eq: Edge Pruning Topology Optimization Training},
\end{equation}
where ${\Delta}s$ is the edge score, and $T_s(t)$ is the dynamic threshold updated by the following rule:
\begin{equation}
{T_{s}(t)}=\begin{cases}
 {T_{s}^{init}}& \text{ if } t=0 \\
 {T_{s}(t-1)} - \frac{T_{s}^{init} - {T_{s}^{end}}}{\alpha} & \text{ } otherwise
\end{cases}
\label{eq:decay}.
\end{equation}
The $T_{s}^{init}$, $T_{s}^{end}$, and ${\alpha}$ represent the pre-designed initial threshold, end threshold, and update step. When a new best accuracy is achieved, the value of $t$ is increased by 1. This decaying threshold speeds up model convergence and is inspired by the learning rate decay technique~\cite{learningrate}.

\subsubsection*{Weight Optimization}

\begin{equation}
{\Delta}w=\begin{cases}
 {\Delta}w& \text{ if } |{\Delta}w|\ge T_w \\
 0& \text{ } otherwise
\end{cases}
\label{eq:weight optimization},
\end{equation}
where ${\Delta}w$ represents the weight updates, and $T_w$ denotes the pre-designed threshold constant used to ascertain if a particular cell requires programming. When $T_w$ equals zero, each cell is programmed during every training step, corresponding to weight optimization with free updates. By increasing the value of $T_w$, the frequency of cell programming decreases, and weight optimization with limited updates is obtained when its programming count is similar to the edge pruning topology optimization.

\subsection*{Details of the experimental Neural Networks}

When performing randomly weighted neural networks, the analogue resistive memory in-memory computing chip accelerates the most computationally expensive vector-matrix multiplications, while the remaining maxpooling, normalization, and activation operations are handled by the Xilinx SoC. The training hyperparameters are detailed in the Supplementary Information.

\subsubsection*{CNN Backbone}

As shown in Fig. \ref{fig3}a, the 4-layer CNN consists of 2 convolutional (C) and 2 fully connected (F) layers. The C1 layer has a kernel size of 64×1×3×3 (out channel × in channel × weight × height) and produces feature maps of size 64×12×12 (out channel × weight × height). The C2 layer has a kernel size of 16×64×3×3 and generates feature maps of size 16×10×10. These maps are then sub-sampled by a maxpooling layer with a 2×2 kernel and stride of 2. Subsequently, the 16×5×5 maps are flattened into a 400-element vector fed into the F1 layer. The F1 layer has a weight size of 128×400 and reduces the input to a 128-element vector. Finally, the vector is passed through the F2 layer with 10×128 weights to obtain the 10 probability outputs that determine the label of the tested image. 

\subsubsection*{CRNN Backbone}

The Spoken Digits are first converted into frequency-domain using the mel-frequency cepstral coefficients (MFCC) method~\cite{choi2017convolutional}, generating a set of cepstral feature maps that can be effectively recognized and classified. As shown in Fig. \ref{fig4}a, these maps with a size of 23×15 are then fed into a 5-layer CRNN consisting of 2 convolutional layers, 1 recurrent layer (R), and 2 fully connected layers. The C1 layer has a kernel size of 64×1×3×2, followed by a maxpooling layer that produces feature maps of size 64×11×14. The C2 layer has a kernel size of 32×64×3×2, followed by a maxpooling layer that generates feature maps of size 32×2×8. These maps are then sent into the R1 layer:
\begin{equation}
{h(t)}=\begin{cases}
 0& \text{ if } (t=0) \\
 {tanh(w_{ih}x_t + w_{hh}h(t-1))}& \text{ } otherwise
\end{cases}
\label{eq:RNN1},
\end{equation}
where ${t}$, ${h(t)}$, $w_{ih}$, and $w_{hh}$ denote the time step, hidden state, input hidden weights (128×32), and recurrent hidden weights (128×128). The time step increases as the model iterates, and when it reaches 4, the final output of the R1 layer is obtained by:
\begin{equation}
{recurrent\;\;output = \frac{\sum_{t=1}^{4 }{h(t)}}{4}}
\label{eq:RNN2}.
\end{equation}
Subsequently, the recurrent output with 128 elements is fed into the F1 (256×128) and F2 (10×256) layers to obtain the 10 probability outputs that determine the label of the input Spoken Digits.

\subsubsection*{U-Net Backbone}

As shown in Fig. \ref{fig5}a, the U-Net architecture consists of two main parts: the contracting path (down-sampling layer, D) and the expansive path (up-sampling layer, U). The down-sampling layer applies a sequence of two 3×3 convolutions, each followed by a rectified linear unit (ReLU) activation and 2×2 max-pooling. This process enables the network to capture relevant features from the input image, allowing it to identify critical structures and patterns. The up-sampling layer, on the other hand, involves a bilinear interpolation layer, a concatenation with the corresponding cropped feature map, and two 3×3 convolutions, each followed by a ReLU activation. This process facilitates precise localization and segmentation of objects within the image.

During simulations, the eyeball blood vessel pictures with dimensions of 584×565 pixels are first split into 26,400 96×96 overlapped patches and then randomly sampled 6,400 patches as the training set while the remaining 20,000 patches are used as the test set. These patches are fed into the input convolutional layer with a kernel size of 32×1×3×3 for the initial feature extraction. Subsequently, these feature maps are sent into four down-sampling layers (D1-D4) and four up-sampling layers (U1-U4) for blood vessel segmentation. The input channels for D1 to D4 kernels are 32, 64, 128, and 256, while those for U1 to U4 are 256, 128, 64, and 32. The output convolutional layer with a kernel size of 2×32×1×1 transforms the U4's results into well-processed patches. Finally, the overall segmented image is generated by concatenating these individual segments.

\subsubsection*{Image Segmentation Indicators}
To demonstrate the performance of U-Net in the DRIVE image segmentation task, several indicators are evaluated, including True Positive Rate (TPR), False Positive Rate (FPR), Recall, Precision, F1-score (F1), and Accuracy, as follows:
\begin{equation}TPR = Recall = \frac{TP}{TP+FN}\label{eq:TPR},\end{equation}
\begin{equation}FPR = \frac{FP}{TN+FP}\label{eq:FPR},\end{equation}
\begin{equation}Precision = \frac{TP}{TP+FP}\label{eq:Precision},\end{equation}
\begin{equation}F1 = 2 \times \frac{Precision \times Recall}{Precision + Recall}\label{eq:F1}.\end{equation}
\begin{equation}Accuracy = \frac{TP+TN}{TP+FP+TN+FN}\label{eq:Accuracy},\end{equation}
Here, recall, precision, F1-score, and accuracy are widely used performance metrics for evaluating image segmentation tasks. Recall evaluates the U-Net's ability to identify all relevant instances, precision measures the correctness of positive predictions, F1-score provides a balanced value ranging from 0 to 1, with higher values representing better classification performance, and the accuracy indicates the proportion of correct predictions made out of the total number of predictions. These metrics help assess classification performance accurately.

\section*{Acknowledgements}
This research is supported by the National Key R\&D Program of China (Grant No. 2022YFB3608300), the National Natural Science Foundation of China (Grant Nos. 62122004, 61874138, 61888102, 61821091), the Strategic Priority Research Program of the Chinese Academy of Sciences (Grant No. XDB44000000), Beijing Natural Science Foundation (Grant No. Z210006), and Hong Kong Research Grant Council (Grant Nos. 27206321, 17205922, 17212923). This research is also partially supported by ACCESS – AI Chip Center for Emerging Smart Systems, sponsored by Innovation and Technology Fund (ITF), Hong Kong SAR.

\section*{Competing Interests}
The authors declare no competing interests.

\bibliography{reference}

\end{document}